\documentclass[prd,superscriptaddress,showpacs,nofootinbib,preprintnumbers]{revtex4}

\usepackage{amsmath}
\usepackage{amsfonts}
\usepackage{graphicx}
\usepackage{dcolumn}
\def\hs{\qquad} 
\def\beq{\begin{eqnarray}} 
\def\eeq{\end{eqnarray}} 
\def\ap{\left.} 
\def\at{\left(} 
\def\aq{\left[} 
\def\ag{\left\{} 
\def\cp{\right.} 
\def\ct{\right)} 
\def\cq{\right]} 
\def\lap{\Delta\,} 
\def\ii{\infty}
\def\segue{\qquad\Longrightarrow\qquad} 
\def\al{\alpha}
\def\be{\beta}
\def\ga{\gamma}
\def\de{\delta}
\def\ep{\varepsilon}
\def\ze{\zeta}

\def\ka{\kappa}
\def\la{\lambda}

\def\si{\sigma}

\def\ph{\varphi}

\def\Ga{\Gamma}

\def\La{\Lambda}

\newcommand{\bea}{\begin{eqnarray}}
\newcommand{\eea}{\end{eqnarray}}
\newcommand{\beaa}{\begin{eqnarray*}}
\newcommand{\eeaa}{\end{eqnarray*}}

\def\nn{\nonumber}
\newcommand{\e}{{\rm e}}

\def\h{\tilde h}
\def\txi{\tilde\xi}
\def\B{\tilde B}
\def\C{\tilde C}
\def\R{R_0}
\def\g{\hat g}
\def\V{\tilde V}

\begin{document}
\tolerance=5000

\title{One-loop effective action for non-local modified Gauss-Bonnet
gravity \\ in de Sitter space}

\author{
 Guido Cognola$\,^{(a)}$\footnote{cognola@science.unitn.it},
 Emilio Elizalde$\,^{(b)}$\footnote{elizalde@ieec.uab.es},
 Shin'ichi Nojiri$\,^{(c)}$\footnote{nojiri@nda.ac.jp,
 snojiri@yukawa.kyoto-u.ac.jp},\\
 Sergei D.~Odintsov$\,^{(b,d)}$\footnote{odintsov@ieec.uab.es also at
Center of Theor.Phys., TSPU,
 Tomsk} and
 Sergio Zerbini$\,^{(a)}$\footnote{zerbini@science.unitn.it}
}

\affiliation{
$^{(a)}$ Dipartimento di Fisica, Universit\`a di Trento \\
and Istituto Nazionale di Fisica Nucleare \\
Gruppo Collegato di Trento, Italia\\
\medskip
$^{(b)}$ Consejo Superior de Investigaciones Cient\'{\i}ficas
(ICE/CSIC) \, and \\ Institut d'Estudis Espacials de Catalunya
(IEEC) \\
Campus UAB, Facultat Ci\`encies, Torre C5-Par-2a pl \\ E-08193 Bellaterra
(Barcelona) Spain\\
\medskip
$^{(c)}$ Department of Physics, Nagoya University, Nagoya
464-8602. Japan, \\
\medskip
$^{(d)}$ ICREA, Barcelona, Spain \, and \,
ICE (CSIC-IEEC) \\
Campus UAB, Facultat Ci\`encies, Torre C5-Par-2a pl \\ E-08193 Bellaterra
(Barcelona) Spain\\
}

\begin{abstract}
We discuss the classical and quantum properties of non-local modified
Gauss-Bonnet gravity in de Sitter space, using its equivalent
representation via string-inspired local scalar-Gauss-Bonnet gravity
with a scalar potential. A classical, multiply de Sitter universe
solution is found where one of the de Sitter phases corresponds
to the primordial inflationary epoch, while the other de Sitter space
solution---the one with the smallest Hubble rate---describes the late-time
acceleration of our universe. A Chameleon scenario for the theory under
investigation is developed, and it is successfully used to show that the
theory complies with gravitational tests. An explicit expression
for the one-loop effective action for this non-local modified
Gauss-Bonnet gravity in the de Sitter space is obtained. It is argued
that this effective action might be an important step towards the 
solution of the cosmological constant problem.

\end{abstract}


\maketitle

\section{Introduction \label{S:intro}}

The discovery of the accelerated expansion of the late-time universe
was the starting point of a wide spectrum of different theoretical
constructions which aim is to provide a reasonable explanation of this acceleration.
The simplest qualitative possibility for such construction is to
consider a gravitational modification of General Relativity (GR), as compared with
the introduction of extra, exotic dark components of the energy in an ordinary GR scheme.
There are a number of different candidates which qualify for gravitational alternatives of
dark energy (a general review of those theories can be found in \cite{review}).
Recently, as a new key proposal for dark energy, {\it non-local} gravitational
theories have been considered \cite{n1,n2,Capozziello:2008gu}.
It has been demonstrated that some versions of these non-local gravities, which depend
on the curvature and its (inverse) derivatives only, are definitely
able to pass the Solar System tests \cite{n1,n2}.
A scalar-tensor representation for such theories has been developed too \cite{n2},
making explicit the connection with local gravities.
Furthermore, the possibility to construct a unified description of the early-time
inflation epoch with the late-time acceleration period becomes quite natural in
such theories \cite{n2} (for related, string-inspired non-local theories with scalars
as dark-energy models, see also \cite{n3} and the references therein).

In this paper we study classical and quantum aspects of the
non-local Gauss-Bonnet gravities introduced by some of the present authors
in \cite{Capozziello:2008gu}, in the de Sitter space.
Their massive and scalar-potential versions are proposed, and their relation with
local string-inspired scalar-Gauss-Bonnet gravity will be investigated.
Using the above equivalence, the corresponding de Sitter solution with a constant scalar
field will be explicitly constructed.
The Chameleon scenario for this theory will be investigated too, and it will be shown
that the theory under consideration passes the local tests (Newton's law, absence 
of instabilities). The one-loop effective action for the theory under
discussion (again, using its equivalence with the local version) will be explicitly
evaluated on the de Sitter space. The corresponding effective action is then
explicitly obtained by using zeta-regularization, and it is used in the important 
discussion of the induced cosmological constant. Finally, some different version of
massive non-local GB gravity, which may also be presented as a local
multi-scalar-GB theory, is proposed. A classical, multiple de Sitter
solution of this theory is found, where one of the de Sitter points can serve for the
description of the inflationary epoch, while the other de Sitter solution, with
much smaller value of the corresponding Hubble parameter, can be used for the description of
the late-time acceleration period.

\section{Non-local modified Gauss-Bonnet gravity as string-inspired
scalar-Gauss-Bonnet theory \label{S:NonLGB}}

We consider the non-local Gauss-Bonnet (GB) model introduced
in \cite{Capozziello:2008gu},
\beq
\label{nlGB1}
S=\int d^4x\sqrt{-g}\left(\frac{R}{2\kappa^2}
 -\frac{\kappa^2}{2a}{\cal G}\Box^{-1}{\cal G}\right)
+S_m\ ,
\eeq
where $S_m$ is the matter action,
$R$ the scalar curvature, ${\cal G}$
the Gauss-Bonnet invariant, and $\Box$ the d'Alembertian operator in the metric $g_{ij}$,
with determinant $g$.
Finally, $\ka$ is related to the Newton constant $G$ by
$\ka^2=8\pi G/c^3$.
By introducing the scalar field $\phi$,
one can rewrite the action (\ref{nlGB1}) in a local form, namely
\beq
\label{nlGB3}
S=\int d^4 x\sqrt{-g}\left(\frac{R}{2\kappa^2}
 -\frac{a}{2\kappa^2}\,g^{ij}\partial_i\phi\partial_j\phi
 +\phi {\cal G}\right)+S_m\ .
\eeq
In fact, one of the field equations of the latter action gives
$\phi = - \frac{\kappa^2}{a} \Box^{-1} {\cal G}$.
By substituting this expression into (\ref{nlGB3}), one
obtains (\ref{nlGB1}). Note that the action (\ref{nlGB3}) corresponds
to string-inspired scalar GB gravity, which was
proposed as a dark energy model in Ref.~\cite{scalar-Gauss-Bonnet,Ito:2009nk}.

For the sake of generality, we add a potential $V(\phi)$
to the Lagrangian density in (\ref{nlGB3}), that is
\beq
\label{nlGBV1}
S=\int d^4 x \left(\frac{R}{2\kappa^2}
 -\frac{a}{2\kappa^2}\,g^{ij}\partial_i\phi\partial_j\phi
 -V(\phi) + \phi {\cal G}\right)+S_m\ .
\eeq
If we eliminate $\phi$ by using the field equation
$\Box\phi = V'(\phi)-\frac{\kappa^2}{a} {\cal G}$,
we obtain a rather untractable non-local theory
(here and in the following $V',V''$ mean derivatives with 
respect to the argument).
There is however a simple case, that is
\beq
\label{nlGBV1b}
V(\phi) = \frac{a}{2\kappa^2}m^2 \phi^2,
\eeq
where a mass term is present, which, typically, may be thought of as a
non-perturbative string correction. In case of (\ref{nlGBV1b}),
by eliminating the scalar field $\phi$, we obtain
\beq
\label{nlGB1b}
S=\int d^4x \sqrt{-g}\left(\frac{R}{2\kappa^2}
 -\frac{\kappa^2}{2a}{\cal G}
 \left(\Box-m^2\right)^{-1}{\cal G}\right)+S_m\ .
\eeq

Coming back to action (\ref{nlGBV1}), by assuming the metric to be the FRW one,
with flat spatial section, and $\phi$ to depend on the cosmological time only,
the equations of motion---neglecting for simplicity the contribution due 
to matter---read
\beq
\label{nlGBV2}
0=-\frac{3H^2}{\kappa^2}+\frac{a}{2\kappa^2}{\dot\phi}^2
+ V(\phi)-24\dot\phi H^3\ ,\quad
0=-\frac{a}{\kappa^2}\left(\ddot\phi+3H\dot\phi\right)
 -V'(\phi)+ 24\left(\dot H H^2 + H^4 \right)\ .
\eeq
As usual, $H=H(t)$ represents the Hubble parameter and
the dot, as in $\dot H,\dot\phi$, means derivative with respect to the
cosmological time. If we further restrict to de Sitter space,
that is, if we take $H=H_0$ and $\phi=\phi_0$ to be constant, we obtain
\beq
\label{nlGBV3}
V(\phi_0)=\frac{6H_0^2}{\kappa^2}\ ,\quad
V'(\phi_0)=24H_0^4 \ .
\eeq
Thus, we see that there is always a de Sitter solution if
the potential satisfies the condition (\ref{nlGBV3})
for a specific value of $\phi$.

For example, if $V$ is a mass term as in Eq.~(\ref{nlGBV1b}), then
from (\ref{nlGBV3}) we get
\beq
H_0^2=\at\frac{m^2}{96\ka^2}\ct^{\frac13}\, ,\hs\hs
\phi_0=\at\frac{3}{2\ka^4m^2}\ct^{\frac13}\,.
\eeq

As a second---non trivial---example we consider the potential
\beq
\label{GBV4}
V=V_0 \e^{q \phi}\ ,
\eeq
with positive constants $V_0$ and $q_0$. The model has
a de Sitter solution, where
\beq
\label{GBV5}
H_0^2 = \frac{q}{8\kappa^2}\ ,\quad
\phi_0 = \frac{1}{q}\ln \left(\frac{3q}{8\kappa^4 V_0}\right)\ .
\eeq
In \cite{Capozziello:2008gu} it has been shown that a classical de Sitter
solution exists in the absence of the potential, too, but in such case
the background field $\phi$ is a time-dependent function. Extending this
formulation one can construct de Sitter solution with time-dependent
scalars also in the presence of the potential. However, such solutions
will not be discussed here, due to fact that we are primarily interesting in the
quantum properties of non-local modified GB gravity on the de Sitter
background with constant scalars.

\section{Chameleon scenario in non-local modified Gauss-Bonnet gravity}

Many dark energy models generically include propagating
scalar modes, which might render a large correction to Newton's law,
if the scalar field couples with usual matter. In order to avoid this problem,
a scenario called the Chameleon mechanism has been proposed \cite{Chame,Mota:2003tc}.
In this scenario, the mass of the scalar mode
becomes large due to the coupling with matter or the scalar curvature in the Solar System and/or
on the Earth. Since the range of the force mediated by the scalar field is given by the Compton
length, if the mass is large enough--and therefore the Compton length is
short enough--the correction to Newton's law becomes very small and it cannot be observed.
The Chameleon mechanism has been used to obtain realistic models of
$F(R)$-gravity \cite{Hu:2007nk,acc,acc1}.
In this section, we are going to consider the variant of the Chameleon mechanism which
is generated by the coupling of the
scalar field with the GB invariant in the action. A mechanism of this kind
has been proposed in \cite{Ito:2009nk}.
As we have shown, the non-local GB theory can be rewritten as a
model with a scalar field coupled with
the Gauss-Bonnet invariant. Then, the Chameleon mechanism could work in a theory of this class,
which we will investigate in the present section.

For the action (\ref{nlGBV1}) the mass of the field is given by
\begin{equation}
\label{nlGBChame1}
m_\phi^2 = \frac{\kappa^2 V''(\phi_e)}{a}\ .
\end{equation}
Here $\phi_e$ is the background value of $\phi$ in a local region, like on the Earth
or the entire Solar System.
In the relevant region in which we are investigating the possible corrections to Newton's law,
the curvature and
the background scalar field could be almost constant. Then the equation given by the variation
of the action (\ref{nlGBV1})
with respect to $\phi$ takes the following form:
\begin{equation}
\label{nlGBChame2}
V'(\phi_e) = \mathcal{G}_e\ .
\end{equation}
Here, $\mathcal{G}_e$ is the background value of $\mathcal{G}$. For example, on the Earth,
we find that $\mathcal{G}_e \sim 10^{-71}\,\mathrm{eV}^4$.

For the model (\ref{nlGBV1b}) the mass (\ref{nlGBChame1}) is given by
\begin{equation}
\label{nlGBChame3}
m_\phi^2 = \frac{m^2 \kappa^2}{a}\ .
\end{equation}
Since $H_0\sim 10^{-33}\,\mathrm{eV}$ and $1/\kappa \sim 10^{28}\,\mathrm{eV}$,
by using (\ref{nlGBV3})
we find
\begin{equation}
\label{nlGBChame4}
m_\phi^2 \sim \frac{10^{-308}}{a}\,\mathrm{eV}^2\ .
\end{equation}
In order that the Compton length could be $1\,\mu \mathrm{m}$,
that is, $m_\phi\sim 1\,\mathrm{eV}$, we find that $a\sim 10^{-308}$,
which is very small but, anyway, the Chameleon mechanism could perfectly work.

For the model (\ref{GBV4}) Eq.~(\ref{nlGBChame2}) has the form:
\begin{equation}
\label{nlGBChame5}
q V_0 \e^{q\phi_e} = \mathcal{G}_e\ ,
\end{equation}
and the mass (\ref{nlGBChame3}) is given by
\begin{equation}
\label{nlGBChame6}
m_\phi^2 = \frac{ \kappa^2 q V_0 \e^{q\phi_e}}{a}\ .
\end{equation}
Then, by using (\ref{GBV5}), we find
\begin{equation}
\label{nlGBChame7}
m_\phi^2 \sim \frac{10^{-248}}{a}\,\mathrm{eV}^2\ .
\end{equation}
Thus, when $a\sim 10^{-248}$ the Compton length could be $1\,\mu \mathrm{m}$.

For a different example, we can now consider the following model,
\begin{equation}
\label{RR1}
V(\phi)= \frac{V_0}{\left(\phi - \alpha\right)\left(\phi^2 + \beta^2\right)}\ .
\end{equation}
Here $V_0$ and $\beta$ are positive constants, and $\alpha$ is a constant.
As long as $\phi>\alpha$, $V(\phi)$ is a
positive and smooth function of $\phi$.
In this case, Eqs.~(\ref{nlGBV2}) have the following form
\begin{equation}
\label{RR2}
\frac{V_0}{\left(\phi_0 - \alpha\right)\left(\phi_0^2 + \beta^2\right)}
 = \frac{3H_0^2}{\kappa^2}\ ,\quad
 - \frac{V_0 \left(3\phi^2 - 2 \alpha \phi_0 + \beta^2 \right)}{\left(\phi_0 - \alpha\right)^2
\left(\phi_0^2 + \beta^2\right)^2} = 24H_0^4 \ .
\end{equation}
By eliminating $H_0$ in the two above equations (\ref{RR2}), we obtain
\begin{equation}
\label{RR3}
3\phi^2 - 2\alpha \phi + \beta^2 + \frac{8}{3}\kappa^4 V_0 = 0\ ,
\end{equation}
which can be solved as follows:
\begin{equation}
\label{RR4}
\phi = \phi_\pm \equiv \frac{\alpha \pm \sqrt{\alpha^2 - 3\beta^2 - 8 \kappa^2 V_0}}{3}\ .
\end{equation}
Since $\phi_+ > \phi_-$, as long as $\alpha<0$, we find $\phi_\pm > \alpha$.
Then, $V(\phi_\pm)$ is surely positive and therefore there are two de Sitter solutions
where the Hubble rate is given by
\begin{equation}
\label{RR6}
H_0^2 = H_\pm^2 \equiv \frac{\kappa^2}{3}\left( \frac{V_0}{\left(\phi_\pm - \alpha\right)
\left(\phi_\pm ^2 + \beta^2\right)} \right)\ .
\end{equation}
The smaller one could be identified with the present acceleratedly expanding universe
and the larger one, with the inflation epoch in the early universe.
We should note that, since there is no singularity in $V(\phi)$ for $\phi_-<\phi<\phi_+$,
the two solutions are connected smoothly and, therefore, the transition from the
inflationary era to the Dark Energy universe is possible, in principle.
On Earth or at the Solar System scale, Eq.~(\ref{nlGBChame2}) acquires the following form:
\begin{equation}
\label{RR7}
- \frac{V_0\left(3\phi_e^2 - 2 \alpha \phi_e + \beta^2 \right)}{\left(\phi_e - \alpha\right)^2
\left(\phi_e^2 + \beta^2\right)^2} = \mathcal{G}_e\ ,
\end{equation}
which could be solved with respect to $\phi_e$. The mass (\ref{nlGBChame3}) is given by
\begin{equation}
\label{RR8}
m_\phi = \frac{ \kappa^2 V_0 \left\{ 16\alpha \phi_e^3
 - 6 \left(\alpha^2 + \beta^2 \right) \phi_e^2
 - 2\beta^2 \left( \alpha^2 + \beta^2 \right) \right\}}{a\left(\phi_e - \alpha\right)^3
\left(\ph_e^2 + \beta^2\right)^3}\ .
\end{equation}
We can conveniently choose the parameters so that $m_\phi^2$ becomes large enough in order
not to give any measurable correction to Newton's law.

We have thus shown that the Chameleon mechanism can actually work even
in the new situation when we deal with a non-local GB theory
which is equivalent to a local scalar-Einstein-GB theory,
where the scalar field couples with the Gauss-Bonnet invariant. There, even though
the non-local GB theory contains a scalar mode, this scalar mode does not
provide any observable correction to the Newtonian law and, thus, a theory of this kind could 
emerge as a perfectly viable theory. Moreover, as we were able to see, the possibility
to unify early-time inflation with late-time acceleration becomes quite natural in this 
context, what is an added bonus worth mentioning.
Note also that the understanding of the equivalence principle in modified
gravity \cite{thomas} may be somehow different from its standard formulation.

\section{One-loop effective action in the non-local Gauss-Bonnet gravity
on de Sitter space \label{S:OneLQ}}

Here we discuss the one-loop quantization (for a review see Ref.~\cite{buch}) of the classical models
we will be dealing with here, on a maximally symmetric space in the Euclidean approach.
One-loop contributions can be important, especially
during the inflationary phase. In any case, as it was shown in \cite{Cognola:2005sg},
their analysis also provides an alternative method to study the stability with respect
to non-homogeneous perturbations around de Sitter
solutions in modified gravitational models, in agreement with \cite{fara, monica}.

We start with the non-local GB-gravity related to the
generalized model in (\ref{nlGBV1}).
For the sake of simplicity, in the present section we will neglect the matter
action $S_m$, since it is irrelevant for our aims, and we shall
use units in which the speed of light $c=1$ and the Newton constant $16\pi G=1$.
Our analysis is going to be very general: we will consider a non-minimal 
interacting term between the scalar and
gravity of the form $f(\phi){\cal G}$, with $f$ being an arbitrary function.
In this way, the model is described by the (Euclidean) action
\beq
\label{Action}
S=\int d^4x\sqrt{g}\aq R+f(\phi){\cal G}
-a\,g^{ij}\,\partial_i\phi\partial_j\phi-V(\phi)\cq\,.
\eeq
When $f$ is a constant, this action reduces to Einstein's gravity minimally
coupled to the scalar field.

In accordance with the background field method, now we consider the
small fluctuations of the fields around the de Sitter manifold
$(\g_{ij},\phi_0)$, of the kind
\beq
\hat R_{ijrs}=\frac{\R}{12}\at\g_{ir}\g_{js}-\g_{is}\g_{jr}\ct\,,
 \hs\hs\R=4\La= \mbox{const}\,,\hs\phi_0= \mbox{const}\,.
\eeq
This is a classical solution if
the potential satisfies the conditions
\beq
V(\phi_0)=\frac{\R}{2}\,,\hs
V'(\phi_0)=\frac{f'(\phi_0)\R^2}{6}\,.
\label{Vconst}\eeq
which are the analog of (\ref{nlGBV3}) written in terms
of the scalar curvature $\R$.
We see that, if $V(\phi)=0$ ,then the Minkowski solution emerges.

For the arbitrary solutions $(g_{ij},\phi)$ of the field equations, we set
\beq
g_{ij}=\g_{ij}+h_{ij}\,,\hs\hs\phi=\phi_0+\ph\,,
\eeq
and perform a Taylor expansion of the action around the de Sitter
manifold, up to second order in the small
perturbations $(h_{ij},\ph)$.
Before we proceed with the expansion, it is convenient
to write the action in order
to take into account the fact that
${\cal G}$ is a topological invariant. Then we observe that
the non minimal interacting term between gravity
and the scalar field, around the background solution,
can be written in the form
\beq
{\cal G}f(\phi)&=&
 {\cal G}f(\phi_0)+{\cal G}\aq f(\phi)-f(\phi_0)\cq
\nn
\\&=&
 {\cal G}f(\phi_0)+{\cal G}_0\aq f(\phi)-f(\phi_0)\cq
 +({\cal G}-{\cal G}_0)\aq f(\phi)-f(\phi_0)\cq\,.
\label{Act2}
\eeq
Here ${\cal G}_0=\R^2/6$ is the value of the Gauss-Bonnet
invariant evaluated on the de Sitter background.
We note that the first term on the right hand side
of the latter equation
does not give contributions to the classical field equations
and can be dropped out, while the second, proportional to
${\cal G}_0$, modifies the scalar potential.
Then, for our aim it is convenient to
write the classical action (\ref{Action}) in the final form
\beq
\label{Action2}
S=\int d^4x\sqrt{g}\aq R
 +\at f(\phi)-f(\phi_0)\ct({\cal G}-{\cal G}_0)
 -a\,g^{ij}\,\partial_i\phi\partial_j\phi-\V(\phi)\cq\,.
\label{s1}\eeq
where we have introduced the effective potential
\beq
\V(\phi)=V(\phi)-{\cal G}_0[f(\phi)-f(\phi_0)]\,.
\eeq
One can check that action (\ref{s1}) is equivalent to the original 
one (\ref{nlGB3}) when $f(\phi)=\phi$.

We are now ready to perform the expansion.
In the following we will use the compact notation
$\V_0=\V(\phi_0)$, $\V'_0=\V'(\phi_0)$, and so on.
After a straightforward computation, along the same lines as for
one-loop $F(R)$-gravity \cite{Cognola:2005de},
one obtains
\beq
S[h]&\sim& \int\:d^4x\,\sqrt{\g}\:
 \aq\R-\V_0-\V'_0\varphi
 +\at\frac{\R}{4}-\frac{\V_0}{2}\ct h
 +{\cal L}_2\cq\,,
\label{AAA3} \eeq
where ${\cal L}_2$ represents the quadratic
contribution in the fluctuation fields $(h_{ij},\ph)$.
Disregarding total derivatives, this reads
\begin{eqnarray}
{\cal L}_2&=&
 \varphi\at a\Delta-\frac{\V''_0}2\ct\,\varphi
 +\frac14\,\varphi\,\aq
 f'_0\R\at\Delta+\frac23\R\ct+2V'_0\cq\,h
\nonumber\\&&\hs
 +\frac3{32}\,h\,\at-\Delta-\frac{2\V_0}3\ct\,h
 +\frac3{32}\,\si\,\at\Delta+\R-2\V_0\ct
 \,\at-\Delta-\frac{\R}3\ct\,\Delta\,\si
\nonumber\\&&\hs
 -\frac14\,f'_0\R\,\varphi\at-\Delta-\frac\R3\ct\Delta\,\si
 -\frac3{16}\,h\,\at-\Delta-\frac{\R}3\ct\Delta\sigma
\nonumber\\&&\hs
 +\txi^i\,\aq\frac14\,\at2\V_0-\R\ct\,
\at-\Delta-\frac{\R}4\ct\cq\,\txi_i
 +\h^{ij}\,\aq\frac14\,\at\Delta-\frac{\R}3+\V_0\ct\cq\,
 \h_{ij}\,.
\end{eqnarray}
Here $\nabla_k$ and $\Delta=\g^{ij}\nabla_i\nabla_j$
represent the covariant derivative and the Laplace operator, respectively,
in the unperturbed metric $\g_{ij}$.
We have also carried out the standard expansion of the
tensor field $h_{ij}$ in irreducible components \cite{frad,buch}, that is
\beq
h_{ij}&=&\h_{ij}+\nabla_i\txi_j+\nabla_j\txi_i
 +\nabla_i\nabla_j\sigma+\frac14\,g_{ij}(h-\lap\sigma)\:,
\label{tt}
\eeq
where $\si$ is the scalar component, while $\txi_i$
and $\h_{ij}$ are the vector and tensor components, 
respectively, with the properties
\beq
\nabla_i\txi^i=0\:,\hs \nabla_i\h^i_j=0\:,\hs
\h^i_i=0\:.
\label{AAA4}
\eeq

As is well known, invariance under diffeomorphisms
renders the operator in the $(h,\si)$ sector non-invertible.
One needs a gauge fixing term and a corresponding ghost compensating term.
We consider the class of gauge conditions
\beq
\chi_k=\nabla_j h_{jk}-\frac{1+\rho}4\,\nabla_k\,h\:,
\nn
\eeq
parameterized by the real parameter $\rho$.
As gauge fixing, we choose the quite general term \cite{buch}
\beq {\cal L}_{gf}=\frac12\,\chi^i\,G_{ij}\,\chi^j\,,\hs\hs
G_{ij}=\ga\,g_{ij}+\beta\,g_{ij}\lap\,, \label{AAA5}
\eeq
where the term proportional to $\ga$ is the one normally
used in Einstein's gravity.
The corresponding ghost Lagrangian reads \cite{buch}
\beq
{\cal L}_{gh}= B^i\,G_{ik}\frac{\de\,\chi^k}{\de\,\ep^j}C^j\,,
\label{AAA6} \eeq where $C_k$ and $B_k$ are the ghost and anti-ghost
vector fields, respectively, while $\de\,\chi^k$ is the variation of
the gauge condition due to an infinitesimal gauge transformation of
the field. It reads
\beq
\de\,h_{ij}=\nabla_i\ep_j+\nabla_j\ep_i\segue
\frac{\de\,\chi^i}{\de\,\ep^j}=g_{ij}\,\lap
+R_{ij}+\frac{1-\rho}{2}\,\nabla_i\nabla_j\,.
\label{AAA7} \eeq
Neglecting total derivatives one has
\beq {\cal L}_{gh}=B^i\,\at\ga\,H_{ij}+\beta\,\lap\,H_{ij}\ct\,C^j\,,
\label{AAA8}
\eeq
where we have set
\beq
H_{ij}=g_{ij}\at\lap+\frac{R_0}{4}\ct+\frac{1-\rho}{2}\,\nabla_i\nabla_j\,.
\label{AAA9}
\eeq
In terms of irreducible components, one finally obtains
\bea
{\cal L}_{gf} &=&\frac{\ga}2\aq\xi^k\,\at\lap_1+\frac{R_0}4\ct^2\,\xi_k
 +\frac{3\rho}{8}\,h\,\at\lap_0+\frac{R_0}3\ct\,\lap_0\,\si
\cp\nn\\&&\hs\ap
 -\frac{\rho^2}{16}\,h\,\lap_0\,h
-\frac{9}{16}\,\si\,\at\lap_0+\frac{R_0}3\ct^2\,\lap_0\,\si
\cq
\nn\\&&
+\frac{\beta}2\aq\xi^k\,\at\lap_1+\frac{R_0}4\ct^2\,\lap_1\xi_k
 +\frac{3\rho}{8}\,h\,\at\lap_0+\frac{R}4\ct\at\lap_0+\frac{R}3\ct\,
\lap_0 \si
\cp\nn\\&&\hs\ap
 -\frac{\rho^2}{16}\,h\,\at\lap_0+\frac{R_0}4\ct\,\lap_0 h
 -\frac{9}{16}\,\si\,\at\lap_0+\frac{R_0}4\ct\at\lap_0+
\frac{R_0}3\ct^2\,\lap_0 \si \cq\,, \label{AAA10}
\eeq
\beq
{\cal L}_{gh} &=&\ga\aq\hat B^i\at\lap_1+\frac{R_0}{4}\ct\hat C^j
+\frac{\rho-3}{2}\,b\,\at\lap_0-\frac{R_0}{\rho-3}\ct\,\lap_0 c\cq
\nn\\&&\hs +\beta\aq\hat
B^i\,\at\lap_1+\frac{R_0}{4}\ct\,\lap_1\,\hat C^j \cp
\nn \\
 &&\hs\hs\ap +\frac{\rho-3}{2}\,b\,\at\lap_0+\frac{R_0}{4}\ct
 \at\lap_0-\frac{R_0}{\rho-3}\ct\,\lap_0 c\cq\,,
\eea
where ghost irreducible components are defined by
\beq
C_k&=&\C_k+\nabla_k c\,,\hs\hs \nabla_k\C^k=0\,,
\nn\\
B_k&=&\B_k+\nabla_k b\,,\hs\hs \nabla_k\B^k=0\,.
\label{AAA11}
\eeq
In order to compute the one-loop contributions to
the effective action we have to consider the path integral for the
bilinear part
\beq
{\cal L}= {\cal L}_2+\,{\cal L}_{gf}+{\cal L}_{gh} \label{AAA12}
\eeq
of the total Lagrangian, and take into
account the Jacobian due to the change of variables with respect to
the original ones. In this way, the Euclidean one-loop partition function reads \cite{frad,buch}
\beq
Z^{(1)}&=&\at\det G_{ij}\ct^{-1/2}\,\int\,D[h_{ij}]D[C_k]D[B^k]\:
\exp\,\at-\int\,d^4x\,\sqrt{\g}\,{\cal L}\ct
\nn\\
&=&\at\det G_{ij}\ct^{-1/2}\,\det J_1^{-1}\,\det J_2^{1/2}\,
\nn\\
&&\times \int\,D[h]D[\h_{ij}]D[\txi^j]D[\si] D[\C_k]D[\B^k]D[c]D[b]
\:\exp\, \at-\int\,d^4x\,\sqrt{\g}\,{\cal L}\ct\,,
\eeq
where $J_1$ and $J_2$ are the Jacobians coming from the
change of variables in the ghost and tensor sectors, respectively.
They read \cite{buch}
\beq
J_1=\lap_0\,,\hs\hs
J_2=\at-\lap_1-\frac{\R}{4}
\ct\at-\lap_0-\frac{\R}{3}\ct\,\lap_0\,.
\label{AAA13}
\eeq
Finally, the determinant of the operator $G_{ij}$,
acting on vectors assumes, in our gauge, the form
\beq
\det G_{ij}=\mbox{const}\,
 \det\at\lap_1+\frac{\ga}{\beta}\ct\,
 \det\at\lap_0+\frac{R_0}4+\frac{\ga}{\beta}\ct\,,
\label{AAA14}
\eeq
while it is trivial in the simplest case $\beta=0$.
By $\lap_n$ ($n=0,1,2$) we indicate the Laplacians acting on
scalar, vector, and transverse tensor fields, respectively.

Now, a straightforward computation, disregarding
zero gravity modes and the multiplicative anomaly as well (\cite{Elizalde:1997nd})),
leads to the expression of the one-loop contribution
$Z^{(1)}(\ga,\be,\rho)$ to the Euclidean partition function.
It is a quite involved expression, which depends on
the gauge parameters, and for this reason we will only write it explicitly
in the Landau gauge corresponding to the choice
$\ga=\ii,\be=0,\rho=1$. On-shell, $Z^{(1)}_{(\ga,\be,\rho)}$
is independent of the gauge and it is compatible with
a similar expression obtained in \cite{frad} for
Einstein's theory with a cosmological constant
$\La_0=\La=\R/4$, but in presence of a scalar field.
In fact, we get
\beq
Z^{(1)}_{\rm on-shell}\equiv e^{-\Ga^{(1)}_{\rm on-shell}}=
 \aq\frac{\det\at-\lap_1-\La\ct}
 {\det\at\,-\lap_2+\frac23\,\La\ct}\cq^{1/2}\,
 \aq\det\at\,-\lap_0+\frac{\V''_0}2\ct\cq^{-1/2}\,.
\label{PFonshell}
\eeq
The latter term is due to the scalar field, but it also
depends on the coupling with the Gauss-Bonnet invariant.
It has to be noted that the on-shell partition function
is obtained by imposing conditions (\ref{Vconst}) and ${\cal G}=\R/6$.
The latter condition in the expression of the
gauge-dependent one-loop partition function
is equivalent to dropping all terms proportional to $f'_0$.

Now, we explicitly write the off-shell partition function
in Landau's gauge. It reads
\beq
Z^{(1)}_{(\ii,0,1)}&=&
 \aq\frac{\det\at-\lap_1-\La\ct}
 {\det\at-\lap_2+\frac83\,\La-V_0\ct}\cq^{1/2}\,
 \det\at-\lap_0-\frac{\R}2\ct\,
\nn\\&&\hs\hs\times
 \aq\det\at-\lap_0-\frac\R{12}\,q_1\ct\,
 \det\at-\lap_0-\frac\R{12}\,q_2\ct\,
 \det\at-\lap_0-\frac\R{12}\,q_3\ct\,\cq^{-1/2}
\,.\label{PF}
\eeq
The quantities $q_1$, $q_2$, $q_3$, which depend in general on $\La$,
are the roots of the third-order algebraic equation
\beq
q^3+c_2q^2+c_1q+c_0=0\,,
\label{C3}\eeq
where
\beq
c_0&=&\frac{
4(f'_0)^2\R^4-24f'_0\R^2\V'_0+36(\V'_0)^2-36\V_0\V''_0}
{6a+(f'_0)^2\R^2}\,,
\,,
\nn\\
c_1&=&\frac{2[10(f'_0)^2\R^4-48f'_0\R^2\V'_0
 +36a\R\V_0+54(\V'_0)^2-9\R\V''_0-18\V_0\V''_0]}
{\R[6a+(f'_0)^2\R^2]}\,,
\label{ccc3}\\
c_2&=&\frac{4[7f'_0\R^3+9a(\R+4\V_0)-18f'_0\R\V'_0-9V''_0)]}
{\R[6a+(f'_0)^2\R^2]}\,.
\nn\eeq

The one-loop effective action can now be evaluated by making use of zeta-function
regularization, and actually computing the zeta-functions $\ze(s|L_n)$ related to the
differential-elliptic Laplace-like operators $L_n$ \cite{eli94}.
Using the same notations as in Ref.~\cite{Cognola:2005de},
we have (see the Appendix)
\beq
\Ga^{(1)}_{(\ii,0,1)}&=&
\frac12\,Q_0\at\frac{33}4\ct
+\frac12\,Q_1\at\frac{25}4\ct
-\frac12\,Q_2\at\frac{49}4-\frac{6\La_0}{\La}\ct
\nn\\ &&\hs
-\frac12\,Q_0\at\frac94+q_1\ct
-\frac12\,Q_0\at\frac94+q_2\ct
-\frac12\,Q_0\at\frac94+q_3\ct\,,
\label{Ga1}
\eeq
where
\beq
Q_n(\al)=\ze'\at0|L_n/\mu^2\ct\,,\hs\hs
L_n=-\lap_n-\frac{\R}{12}\,(\al-\al_n)\,,
\eeq
\beq
\ze'(0|L_n)=\lim_{s\to0}\,\frac{d}{ds}\,
\ze(s|L_n)=-\log\det L_n\,,
\eeq
and
$\al_0=9/4$, $\al_1=13/4$, $\al_2=17/4$.

The effective equation for the induced cosmological constant can be obtained by varying the
effective action with respect to $\La$ \cite{frad}.
We would now like to study the role of the Gauss-Bonnet term. For the sake of simplicity,
we choose a scalar potential satisfying the conditions
$\V'_0=0$, $\V''_0=0$. This means that, in the absence of
the ${\cal G}$ term, namely when $f(\phi)=0$, the potential is constant
and the action becomes the Einsteinian one with a cosmological constant
$\La_0=V_0/2$, while, when $f(\phi)=\phi$, there is a coupling of
the form $\phi({\cal G}-{\cal G}_0)$.

We consider separately the two cases $f(\phi)=0$ and $f(\phi)=\phi$.
In the first case we get $q_1=6\La_0/\La$, while $q_2=q_3=0$
and so they do not give a contribution. Thus, we get
\beq
\Ga^{(1)}_{(\ii,0,1)}=
\frac12\,Q_0\at\frac{33}4\ct
+\frac12\,Q_1\at\frac{25}4\ct
-\frac12\,Q_2\at-\frac{15}4+6y\ct
-\frac12\,Q_0\at\frac94+6y\ct\,.
\eeq
Here we have introduced the dimensionless variable
$y=\La_0/\La>0$, so the effective equation for the induced cosmological
constant $\La$ can be obtained by taking the derivative of the effective
action with respect to $y$, since
\beq\label{LaInd}
\frac{\partial\Ga}{\partial y}=
\frac{\partial I}{\partial y}+
\frac{\partial\Ga^{(1)}_{(\ii,0,1)}}{\partial y}=0\,,
\eeq
where $I$ is the classical action on the $S^4$ background,
which reads
\beq
I=\int d^4x\,\sqrt{g}\,(\R-V_0)=
 \frac{96\pi^2}{\La}-\frac{48\pi^2\La_0}{\La^2}
=\frac{48\pi^2}{\La_0}\,(2y-y^2)\,.
\eeq
 From (\ref{BBB1a}) and (\ref{BBB1}) we get
\beq
\frac{\partial Q_n(\al)}{\partial y}&\sim&
 -\frac1y\,\at F_\al(0)
 +\sum_{k=0}^2\,\frac{b_k\al^k}{k!}\ct
 \log\frac{\La_0}{3\mu^2y}\,
 +\frac{d\al}{dy}\,\at \frac{dF_\al(0)}{d\al}+b_1+b_2\al\ct\,
 \log\frac{\La_0}{3\mu^2y}
\nn\\&&\hs\hs
 +\frac{d\al}{dy}\,\at \frac{dF'_\al(0)}{d\al}
 +a_1+\ga b_1+(a_2+\ga b_2)\al+\sum_{k=3}^\ii G(k)\al^{k-1}\ct\,.
\eeq
For the first case, at lower order in $y$, from (\ref{LaInd}) we obtain
the equation
\beq\label{La0}
 0 &\sim& \frac{96\pi^2}{\La_0}\,(1-y)
 +\frac{21.15}{y}-54.60+41.32\,y
\nn\\&&\hs
 +24\log\frac{\La_0}{3\mu^2y}
 -18\,y\,\log\frac{\La_0}{3\mu^2y}
 +O(y^2,y^2\log y)\,.
\eeq
In the second case we also have vector and tensor contributions,
as in the previous one but, in addition, we get three scalar contributions
related with the roots of (\ref{C3}).
Using (\ref{Ga1}) and (\ref{LaInd}), at lowest order in $y$ we obtain
\beq\label{La1}
 0 &\sim& \frac{96\pi^2}{\La_0}\,(1-y)
 +\frac{20.83}{y}-57.11
 +\at43.50-2.93\,\frac{a}{\La_0^2}\ct\,y
\nn\\&&\hs
 +25\log\frac{\La_0}{3\mu^2y}
 -\at15-1.70\,\frac{a}{\La_0^2}\ct\,y\,\log\frac{\La_0}{3\mu^2y}
 +O(y^2,y^2\log y)\,.
\eeq

\begin{figure}
\includegraphics[width=8cm]{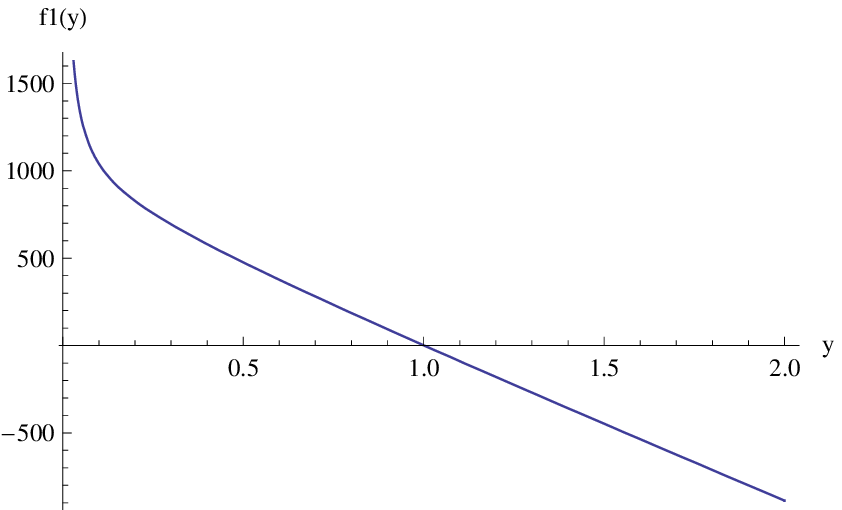}
\hfill
\includegraphics[width=8cm]{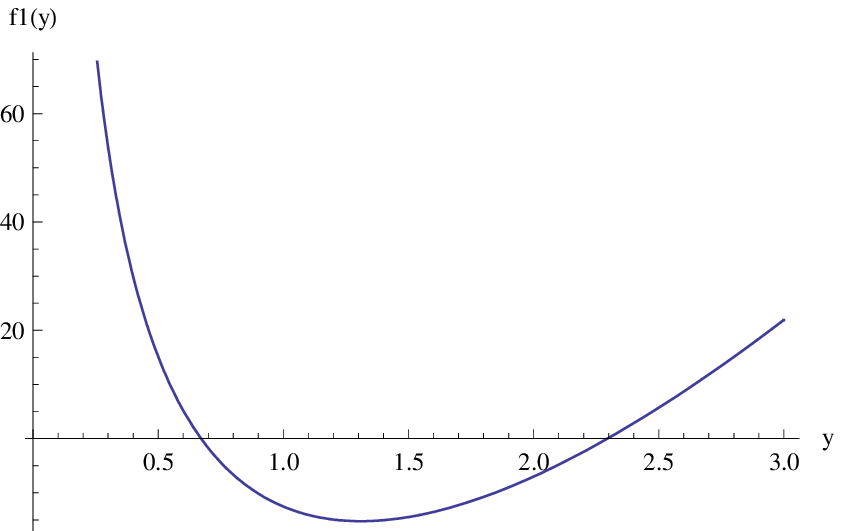}
\caption{{\protect\small Plot of the right-hand-side
of Eq.~(\ref{La0}) vs $y$, for the particular values
of the parameters $\Lambda_0 = \mu^2 =1$ (left figure)
and $\Lambda_0 =\mu^2 =10$ (right figure).}} \label{fig01}
\end{figure}

\begin{figure}
\includegraphics[width=8cm]{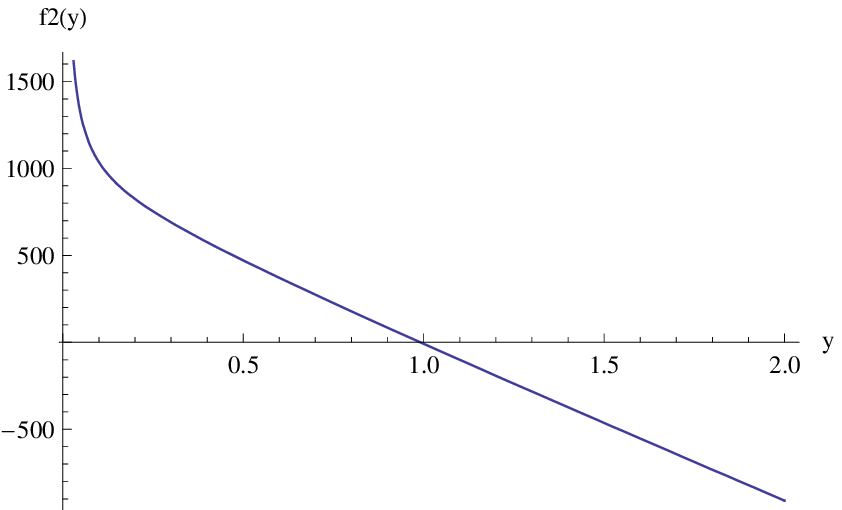}
\hfill
\includegraphics[width=8cm]{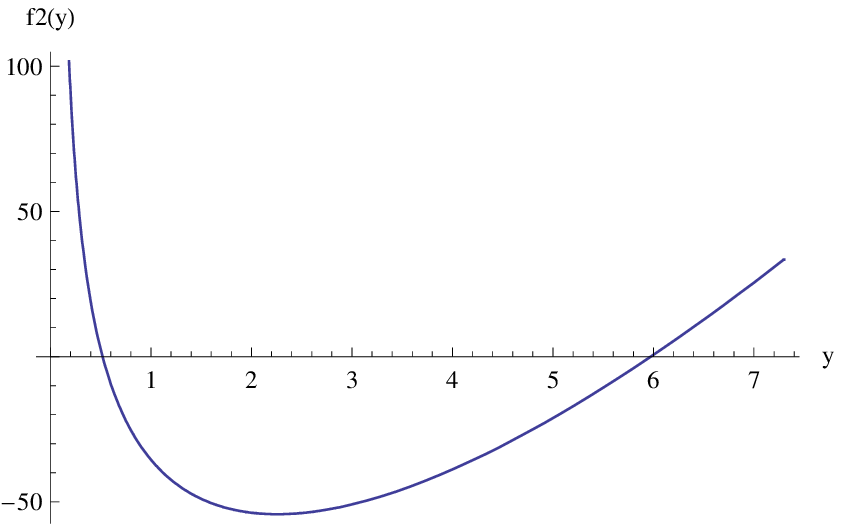}
\caption{{\protect\small Plot of the right-hand-side
of Eq.~(\ref{La1}) vs $y$, for the particular values
of the parameters $\Lambda_0 = \mu^2 =a = 1$ (left figure)
and $\Lambda_0 =\mu^2 =10, a=100$ (right figure).}} \label{fig02}
\end{figure}

In Figs.~(\ref{fig01}) and (\ref{fig02}) we have plotted the right hand sides of
Eqs.~(\ref{La0}) and (\ref{La1}), respectively, as functions of $y$,
for two different sets of values of the parameters. Notice that the zeros of
these functions correspond, in each case, to a zero value of the induced cosmological constant.
Aside from the cases here explicitly depicted, it can be easily seen that in our effective models
the possibilities to get an induced cosmological constant which is exactly zero are reasonably high,
since for values of the parameters lying in wide regions of the domain of expected values,
either one or two roots of the equations exist, yielding the value of
the induced cosmological constant exactly zero.
As is well known \cite{frad}, there is a possible resolution of the the
induced $\Lambda$-term problem coming from the contribution of higher-loop terms.
This is explained in \cite{frad} precisely in the example of an effective action
in pure Einstein gravity for a de Sitter background: one can get a very small
effective $\Lambda$ term irrespective of the tree level cosmological constant.
But an even better possibility is to start from a zero tree level cosmological term,
since quantum corrections will respect this property.
We see that in our model there are good chances to realize this latter situation.
One remark is in order. We got the one-loop effective action for non-local
GB gravity using its classical equivalence with local scalar-GB gravity
and working in terms of such local theory.
It is quite well-known that such classical equivalence may be broken
already at one-loop level.
However, the equivalence is restored on-shell, i.e. using the one-loop
corrected equations of motion. That is precisely the situation in which
the induced cosmological constant has been here discussed.


Now, we come back to the original action (\ref{nlGB3})
with $V(\phi)=0$ and $f(\phi)=\phi$,
which is equivalent to the non-local action (\ref{nlGB1}).
Also for this case there is a de Sitter solution
$\g_{ij},\phi_0$, but $\phi_0$ is not a constant.
This means that the term ${\cal G}f(\phi_0)$ in (\ref{Act2})
cannot be dropped out and, thus, it gives a contribution
to ${\cal L}_2$. In principle, it is possible to
take such contribution into account, but technically
this is quite complicated, since it contains
a lot of independent terms which mix scalar with vector
and tensor components, so that the partition function is
given in terms of the determinant of an involved
$5 \times 5$ matrix of differential operators.

When $V(\phi)=0$ the original model has, however, a
Minkowskian solution. This means that, using (\ref{Ga1})
with $V(\phi)=0$, we can compute the cosmological
constant induced by quantum fluctuations around the Minkowski background,
and this may be interpreted as the spontaneous creation of a de Sitter
universe starting from a flat one, which is a quite interesting feature.

\section{Other classical non-local GB models and their de Sitter
solutions}

Another model giving rise to an interesting non-local action is the following
\beq
\label{nlGBrrr1}
S=\int d^4 x \sqrt{-g}\left[\frac{R}{2\kappa^2}
 - \frac{\kappa^2}{2 a} F({\cal G})
\Box^{-1}F ({\cal G}) \right]\ ,
\eeq
where $a$ is a dimensional constant and $F({\cal G})$ is an adequate function of ${\cal G}$.
By introducing three scalar fields, $\phi$, $\xi$,
and $\eta$, one can rewrite the action (\ref{nlGBrrr1}) under the following form
\beq
\label{nlGBrrr2}
S=\int d^4 x \sqrt{-g} \left(\frac{R}{2\kappa^2}
+ \frac{ a}{2\kappa^2}\partial_\mu \phi \partial^\mu \phi
+ \phi F(\eta)
+ \xi \left(\eta - {\cal G}\right)\right)\ .
\eeq
We may further add a potential $V(\phi)$ to the action
\beq
\label{nlGBrrr2b}
S=\int d^4 x \sqrt{-g} \left(\frac{R}{2\kappa^2}
+ \frac{ a}{2\kappa^2}\partial_\mu \phi \partial^\mu \phi
+ \phi F(\eta) + \xi \left(\eta - {\cal G}\right)
 - V(\phi)\right)\ .
\eeq
In the FRW universe, this action leads to the following equations
\bea
\label{nlGBrrr3}
&& 0 = - \frac{3}{\kappa^2}H^2 - \frac{a}{2\kappa^2}{\dot\phi}^2 + 24 \dot\xi H^3 + V(\phi) \ ,\quad
0 = \frac{a}{\kappa^2} \left(\ddot\phi + 3H\dot\phi\right) + F(\eta) - V'(\phi)\ ,\nonumber \\
&& 0 = \phi F(\eta) + \xi\ ,\quad
0 = \eta - 24 \left(H^4 + \dot H H^2\right)\ .
\eea
However, after adding the potential $V(\phi)$, it is difficult to get
the corresponding non-local action explicitly.

When $V(\phi)=0$, assuming that $\phi = c_\phi t$, $\eta = \eta_0$, $\xi = c_\xi t$, and $H=H_0$,
with constant $c_\phi$, $\eta_0$, $c_\xi$, and $H_0$, the equations in (\ref{nlGBrrr3}) reduce
to the algebraic ones
\beq
\label{nlGBrrr5}
0 = - \frac{3}{\kappa^2}H_0^2 - \frac{a}{2\kappa^2}c_\phi^2 + 24 c_\xi H_0^3\ ,\quad
0 = \frac{3a}{\kappa^2} H_0 c_\phi + F(\eta_0)\ ,\quad
0 = c_\phi F'(\eta_0) + c_\xi\ ,\quad
0 = \eta_0 - 24 H_0^4 \ .
\eeq
We can solve Eqs.~(\ref{nlGBrrr5}) with respect to $c_\phi$, $c_\xi$, and $\eta_0$
as follows: $c_\phi = - \frac{\kappa^2 F\left(24 H_0^4 \right)}{3a H_0}$,
$c_\xi = - \frac{\kappa^2}{3 a H_0}F\left(24 H_0^4 \right)F'\left(24 H_0^4 \right) $,
$\eta_0 = 24 H_0^4$, and we find
\beq
\label{H0}
0 = - \frac{G_0}{8\kappa^2} - \frac{\kappa^2}{18a} F\left(G_0 \right)^2
+ \frac{\kappa^2}{9a} G_0 F\left(G_0 \right) F'\left(G_0 \right)\ .
\eeq
Here $G_0 = 24 H_0^4$.

For example, if we choose
\beq
\label{H1}
F\left({\cal G} \right) = f_0 {\cal G}^2\ ,
\eeq
(\ref{H0}) gives
\beq
\label{H2}
0 = G_0 \left( - \frac{1}{8\kappa^2} + \frac{11 \kappa^2 f_0^2}{18 a} G_0^3 \right) \ ,
\eeq
which has a trivial solution $G_0 = 0$ corresponding to the flat background and
\beq
\label{H3}
G_0^3 = \frac{9a}{44 \kappa^4 f_0^2}\ ,
\eeq
which corresponds to the de Sitter universe.

As another example, one can choose
\beq
\label{H4}
F\left({\cal G} \right)^2 = - g_0 \left({\cal G} + g_1\right) \left( {\cal G} - g_2\right) \ .
\eeq
Here $g_0$, $g_1$, and $g_2$ are positive constants.
In order that $F\left({\cal G} \right)$ can be real, we restrict the value of ${\cal G}$ as
$-g_1 < {\cal G} < g_2$. Then Eq.~(\ref{H0}) yields
\beq
\label{H5}
0 = G_0^2 + \frac{9a}{4 g_0 \kappa^2} G_0 + g_1 g_2\ ,
\eeq
which can be solved as
\beq
\label{H6}
G_0 = G_{0\,\pm} \equiv \frac{1}{2}\left( - \frac{9a}{4 g_0 \kappa^2} \pm
\sqrt{\left(\frac{9a}{4 g_0 \kappa^2}\right)^2 + 4g_1 g_2} \right)\ ,
\eeq
which are positive, provided $a<0$, what we will assume in what follows.
In order that $-g_1 < G_{0\,\pm} < g_2$, we find
\beq
\label{H7}
g_1> - \frac{9a}{4 \kappa^4 g_0}\ ,\quad g_2 > - \frac{9a}{8 \kappa^4 g_0}\ .
\eeq
Then, as long as Eq.~(\ref{H7}) is satisfied, there are two solutions 
describing a de Sitter universe.
By choosing a more general $F({\cal G})$, we may possibly find 
there can be several of these solutions corresponding to de Sitter universes, 
of which the solution with the largest value of $H_0$ could perfectly
correspond to the inflation epoch and the one with the smallest $H_0$ to
the accelerated expansion period of our present universe.
Note that, in the model (\ref{H4}), as far as Eq.~(\ref{H7}) be satisfied,
there is no singularity, say at ${\cal G}=0$. Thus, a smooth transition 
could occur and there exists a true possibility to realize matter
dominance before late-time acceleration.
It is also possible to evaluate the one-loop effective action for the above
theory, too, but the technical details of the calculation are quite 
involved in this case.

\section{Discussion and conclusions}

In summary, we have studied in this paper non-local GB gravity in its
classically-equivalent local scalar-GB form, with a scalar
potential. The classical de Sitter solution was obtained for the model with
one scalar, as well as for the model with several scalars. In the last case,
the multiple de Sitter solution encountered can be used for the unification of the
early-time inflationary epoch with the late-time accelerating one (where an unstable de Sitter
point should be considered for the inflationary epoch). The Chameleon scenario
for non-local GB gravity has been presented and it has been shown that the theory
discussed here satisfies the local gravitational tests (Newton's law).

The one-loop effective action for the theory under investigation has been 
calculated on the de Sitter background. Its explicit representation, in terms 
of a zeta-regularization scheme, gives us the possibility to determine, in all 
precision, the induced cosmological constant. It could well happen in this 
context that, as a result of the quantum corrections, such induced cosmological 
constant could eventually tend to zero, thus providing a natural solution to the
elusive cosmological constant problem.

As a first application of the results obtained in this work, it is quite interesting to
observe that it becomes a rather immediate issue to investigate the stability of our model
around the de Sitter solution. This is in fact necessary, in order to demonstrate that
the de Sitter space is unstable, and most naturally provides a consistent, graceful
exit from the inflationary era. However, that of taking into account infrared quantum
gravity effects and the leading log approximation may eventually be a relevant issue, 
in order to construct the quantum gravity-induced inflationary universe (see e.g. 
\cite{tsamis}). To this aim, it is sufficient to require that the Laplace-like
operators appearing in the on-shell, one-loop partition function
(\ref{PFonshell}) be positive operators \cite{Cognola:2005sg}. As a result, we get
\beq
\V''_0=V''_0-{\cal G}_0\,f''_0
 =V''_0-\frac16\,\R^2\,f''_0 >0\,.
\eeq
For the linear case $f(\phi)=\phi$, this condition reduces
to $V''_0 >0$, and this is certainly satisfied by
the two potentials discussed in Sect.~\ref{S:NonLGB}.

The general, one-loop effective action on the de Sitter space we have found in this
paper can be used as the basis for the search of the non-trivial RG
fixed point in the functional, exact RG approach to string-inspired, scalar-GB
gravity, in the same way as in scalar-Einstein/gauged SG theories \cite{machado},
with account to higher-derivative invariants \cite{percacci} (in our case, the GB term).
Finally, the one-loop calculation performed here is central to the subsequent investigation
of the cosmological perturbation theory in non-local gravities.

\section*{\bf Acknowledgments}

We thank the referee of a previous version of this manuscript for 
remarks and criticisms that definitely contributed to its improvement. 
One of the authors (G.~Cognola) acknowledges the support received from
the European Science Foundation (ESF) for the activity entitled
``New Trends and Applications of the Casimir Effect''
(Exchange Grant 2262) for the period March-April 2009, during
which the present work has been completed.
He his also grateful to the Institute ICE/CSIC in
Barcelona for the kind hospitality received during
that period. This paper is, in part, also an outcome of the
collaboration program INFN (Italy) --- DGICYT (Spain). It has been partly
supported by MEC (Spain), projects FIS2006-02842 and
PIE2007-50/023, by AGAUR (Gene\-ra\-litat de Ca\-ta\-lu\-nya),
contract 2005SGR-00790, and by the Global
COE Program of Nagoya University provided by the Japan Society
for the Promotion of Science (G07).

\appendix

\section{Evaluation of the functional determinants\label{AppX}}

We will here make use of zeta function regularization (see, for example, \cite{eli94})
for the evaluation of the functional determinants appearing in the
one-loop effective action, Eq.~(\ref{PF}) computed in the previous sections.
We shall first outline the standard technique based on binomial
expansion, which relates the zeta-functions corresponding to the
operators $\hat A$, with eigenvalues $\hat\la_n>0$ and
$A=\frac{R_0}{12}(\hat A-\al)$, with eigenvalues
$\la_n=\frac{R_0}{12}(\hat\la_n-\al)$, $\al$ being a real constant.
With this choice, $\hat\la_n$ and $\al$ are dimensionless. We assume
to be dealing with a second-order differential operator on a $D$
dimensional compact manifold. Then, by definition, for $\Re s>D/2$
one has \beq \hat\ze(s)&\equiv&\ze(s|\hat A)=\sum_n\hat\la_n^{-s}\,,
\label{sum1}
\\
\ze_\al(s)&\equiv&\ze(s|A)=\sum_n\la_n^{-s}=\at
\frac{R_0}{12}\ct^{-s} \sum_n(\hat\la_n-\al)^{-s}\,,
\label{sum2}\eeq where, as usual, zero eigenvalues have to be
excluded from the sum. In order the binomial expansion in
(\ref{sum2}) to be meaningful, we have to treat separately the several terms
satisfying the condition $|\hat\la_n|\geq|\al|$. So, we write \beq
\ze_\al(s)=\at \frac{R_0}{12}\ct^{-s}\aq
F_\al(s)+\sum_{k=0}^\ii\frac{\al^k\Ga(s+k)\hat G(s+k)}{k!\Ga(s)}\cq
\,, \label{binE}\eeq
where we have set
\beq
F_\al(s)=\sum_{\hat\la_n\leq|\al|;\,\,\hat\la_n\neq\al}\,(\hat\la_n-\al)^{-s}\,,
\hs\hs \hat F(s)=\sum_{\hat\la_n\leq|\al|}\,\hat\la_n^{-s}\,,
\label{BBB1a}\eeq
\beq
\hat G(s)=\sum_{\hat\la_n>|\al|}\,\hat\la_n^{-s} =\hat\ze(s)-\hat
F(s)\,,\hs\hs\hat F(0)-F_\al(0)=N_0\,,
\label{BBB1}
\eeq
$N_0$ being the number of zero-modes.
It needs to be noted that (\ref{binE})
is valid also in the presence of zero-modes or negative eigenvalues
for the operator $A$. In many interesting cases, $F_\al(s)$ and
$\hat F(s)$ are vanishing and thus $\hat G(s)=\hat\ze(s)$.

As is well known, the zeta function may have simple poles on the real
axis for $s\leq D/2$ but it will be always regular at the origin. Of course, the
same analytic structure is also valid for the function $\hat G(s)$.
One has
\beq
\Ga(s)\hat\ze(s)=\sum_{n=0}^\ii\,\frac{\hat K_n}{s+(n-D)/2}
+\hat J(s)\,,
\label{BBB2}
\eeq
$\hat J(s)$ being an
analytic function and $\hat K_n$ the heat-kernel coefficients
which depend on geometrical invariants. In the physical applications we
want to deal with, we have to work with the zeta function and its
derivative at zero; thus, it is convenient to consider the Laurent
expansion around $s=0$ of the functions
\beq
\Ga(s+k)\hat\ze(s+k)&=&\frac{\hat b_k}{s}+\hat a_k+O(s)\,, \\
\Ga(s+k)\hat G(s+k)&=&\frac{b_k}{s}+a_k+O(s)\,,
\label{BBB3}
\eeq
\beq
b_0&=&\hat b_k-\hat F(0)\,,
\hs\hs
a_0=\hat a_0+\ga\hat F(0)\,,\\
b_k&=&\hat b_k=\hat K_{D-2k}\,,
\hs\hs a_k=\hat a_k-\Ga(k)\hat F(k)\,,
\hs\hs 1\leq k\leq\frac{D}2\,,\\
b_k&=&\hat b_k=0\,,
\hs\hs
\hat G(k)=\hat\ze(k)-\hat F(k)\,,
\hs\hs k>\frac{D}2\,.
\label{BBB4}
\eeq
 From previous considerations, one obtains
\beq
\ze_\al(s)&=&\at \frac{R_0}{12}\ct^{-s} \aq\sum_{0\leq k\leq D/2}\,
\at\frac{b_k\al^k}{k!} +s\,\frac{(a_k+\ga b_k)\al^k}{k!}\ct \cp\nn\\
&& \ap\hs\hs\hs +F_\al(s)+s\sum_{k>D/2}\,\frac{\al^k\hat
G(k)}{k}+O(s^2)\cq \,,\label{BBB5} \eeq and finally \beq
\ze_\al(0)&=&F_\al(0)+\sum_{0\leq k\leq D/2}\,\frac{
b_k\al^k}{k!}\,,
\\
\ze'_\al(0)&=&-\ze_\al(0)\,\log\frac{R_0}{12}+\sum_{0\leq k\leq D/2}\,
\frac{(a_k+\ga b_k)\al^k}{k!}
+F_\al'(0)+\sum_{k>D/2}\,\frac{\al^k\hat G(k)}{k}\,,
\label{ze1}\eeq
$\ga$ being the Euler-Mascheroni constant.
If there are negative eigenvalues then
$F'_\al(0)$ has an imaginary part, which points out to an instability of the model.

In the paper we have to deal with Laplace-like operators acting on scalar
and constrained vector and tensor fields in a 4-dimensional de Sitter
space $SO(4)$.
In all such cases, the eigenvalues $\la_n$ and relative degeneracies $g_n$
can be written in the form
\beq
\la_n=\frac{R_0}{12}\at\hat\la_n-\al\ct\,,
\hs g_n=c_1\,\at n+\nu\ct+c_3\,\at n+\nu\ct^3\,,
\hs\hat\la_n=\at n+\nu\ct^2\,,
\label{hlan}\eeq
where $n=0,1,2...$ and
$c_1,c_2,\nu,\al$ depend on the operator in question. In our case,
 we have
\beq
L_0&=&-\lap_0-\frac{R_0}{12}\,q\segue\ag
\begin{array}{ll}
 \nu=\frac32\,,&
 \al=\frac{9}{4}+q\,,\\ \\
 c_1=-\frac{1}{12}\,,& c_3=\frac13\,.
\end{array}\cp
\label{L0PP}
\\
L_1&=&-\lap_1-\frac{R_0}{12}\,q\segue\ag
\begin{array}{ll}
 \nu=\frac52\,,&
 \al=\frac{13}{4}+q\,,\\ \\
 c_1=-\frac{9}{4}\,,& c_3=1\,.
\end{array}\cp
\label{L1PP}
\\
L_2&=&-\lap_2-\frac{R_0}{12}\,q\segue\ag
\begin{array}{ll}
 \nu=\frac72\,,&
 \al=\frac{17}{4}+q\,,\\ \\
 c_1=-\frac{125}{12}\,,& c_3=\frac53\,,
\end{array}\cp
\label{L2PP}
\eeq
where $q$ are dimensionless parameters depending on the
specific choice of $f(R)$.

We note that $\hat\ze(s)$ is related to the well known
Hurwitz function $\ze_H(s,\nu)$ by
\beq
\hat\ze(s)&=&\sum_{n=0}^\ii\,g_n\hat\la_n^{-s}=
\sum_{n=0}^\ii\,\aq c_1\at n+\nu\ct^{2s-1}
+c_3\at n+\nu\ct^{2s-3}\cq
\nn\\
&=&c_1\ze_H\at 2s-1,\nu\ct+
c_3\ze_H\at 2s-3,\nu\ct
\label{BBB6}
\eeq
and
\beq
\hat G(s)&=&c_1\ze_H\at 2s-1,\nu\ct+c_3\ze_H\at 2s-3,\nu\ct-\hat F(s)
\nn\\
&=&c_1\ze_H\at 2s-1,\nu+\hat n\ct+c_3\ze_H\at 2s-3,\nu+\hat n\ct\,,
\label{BBB7} \eeq
$\hat n$ being the number of terms not satisfying
the condition $\hat\la_n>|\al|$.
In order to proceed, we have to
compute the quantities $\hat b_k$ and $\hat a_k$, for $k=0,1,2$. To
this aim, we note that the Hurwitz function has just a simple pole at 1,
more precisely,
\beq
\ze_H(s+1,\nu)=\frac{1}{s}-\psi(\nu)+O(s)\,,
\label{BBB8}
\eeq
$\psi(s)$ being the logarithmic derivative of Euler's gamma
function. After a straightforward computation, we get
\beq
\hat b_0=\hat\ze(0)=c_1\,\ze_H\at-1,\nu\ct+c_3\,\ze_H\at-3,\nu\ct\,,
 \hs \hat b_1=\frac{c_1}2\,,
 \hs \hat b_2=\frac{c_3}2\,,
\label{bk}
\eeq
\beq
\hat a_0&=&\hat\ze'(0)-\ga\hat\ze(0)
\nn\\&=&c_1\,\aq2\ze'_H\at-1,\nu\ct-\ga\ze_H\at-1,\nu\ct\cq
+c_3\,\aq2\ze'_H\at-3,\nu\ct-\ga\ze_H\at-3,\nu\ct\cq\,,
\\
\hat a_1&=&-c_1\,\aq\psi\at\nu\ct+\frac{\ga}{2}\cq
 +c_3\,\ze_H\at-1,\nu\ct\,,
\\
\hat a_2&=&c_1\,\ze_H\at3,\nu\ct
-c_3\,\aq\psi\at\nu\ct+\frac{\ga-1}{2}\cq\,.
\label{ak}
\eeq
Using (\ref{ze1}), we obtain
\beq
Q_n(\al)\equiv\ze'_\al(0|L_n/\mu^2)&=&
\at F_\al(0)+\sum_{k=0}^2\,\frac{b_k\al^k}{k!}\ct
\log\frac{R_0}{12\mu^2}
\nn\\&&\hs\hs
+\sum_{k=0}^2\,\frac{(a_k+\ga b_k)\al^k}{k!}
+F_\al'(0)+\sum_{k=3}^\ii\,\frac{\al^k\hat G(k)}{k}\,.
\label{BBB9}
\eeq
To conclude we would like to remark that
in (\ref{BBB1a}) and (\ref{BBB1})
it is not strictly necessary that $\la_n<\al$. In
principle one can add up an arbitrary number of terms, with
the only restriction that $\la_n\neq\al$, and all expressions
we have derived here will be still valid. This means that
$\hat n$ is allowed to be an {\it arbitrary} number.
In this way, the convergence of the series in (\ref{BBB9})
can be improved at will, which is a very nice feature of the
procedure.



\begin{thebibliography}{99}

\bibitem{review}
S.~Nojiri and S.~D.~Odintsov,
 eConf {\bf C0602061}, 06 (2006)
 [Int.\ J.\ Geom.\ Meth.\ Mod.\ Phys.\ {\bf 4}, 115 (2007)]
 [arXiv:hep-th/0601213].

\bibitem{n1}
S.~Deser and R.~P.~Woodard,
 Phys.\ Rev.\ Lett.\ {\bf 99}, 111301 (2007)
 [arXiv:0706.2151 [astro-ph]];
C.~Deffayet and R.~P.~Woodard,
 arXiv:0904.0961 [gr-qc];
N.~A.~Koshelev,
 arXiv:0809.4927 [gr-qc].

\bibitem{n2}
S.~Nojiri and S.~D.~Odintsov,
 Phys.\ Lett.\ B {\bf 659}, 821 (2008)
 [arXiv:0708.0924 [hep-th]];
S.~Jhingan, S.~Nojiri, S.~D.~Odintsov, M.~Sami, I.~Thongkool and S.~Zerbini,
 Phys.\ Lett.\ B {\bf 663}, 424 (2008)
 [arXiv:0803.2613 [hep-th]];
T.~Koivisto,
 Phys.\ Rev.\ D {\bf 77}, 123513 (2008)
 [arXiv:0803.3399 [gr-qc]];
T.~S.~Koivisto,
 Phys.\ Rev.\ D {\bf 78}, 123505 (2008)
 [arXiv:0807.3778 [gr-qc]].

\bibitem{Capozziello:2008gu}
S.~Capozziello, E.~Elizalde, S.~Nojiri and S.~D.~Odintsov,
Phys.\ Lett.\ B {\bf 671} (2009) 193
[arXiv:0809.1535 [hep-th]].

\bibitem{n3}
I.~Y.~Aref'eva, A.~S.~Koshelev and S.~Y.~Vernov,
 Phys.\ Rev.\ D {\bf 72}, 064017 (2005)
 [arXiv:astro-ph/0507067];
G.~Calcagni, 
theory,'' JHEP {\bf 0605}, 012 (2006) [arXiv:hep-th/0512259];
I.~Y.~Aref'eva, L.~V.~Joukovskaya and S.~Y.~Vernov,
 JHEP {\bf 0707}, 087 (2007)
 [arXiv:hep-th/0701184];
G.~Calcagni and G.~Nardelli,
 arXiv:0904.4245 [hep-th];
G.~Calcagni, M.~Montobbio and G.~Nardelli,
 Phys.\ Rev.\ D {\bf 76} (2007) 126001
 [arXiv:0705.3043 [hep-th]].

\bibitem{scalar-Gauss-Bonnet}
S.~Nojiri, S.~D.~Odintsov and M.~Sasaki,
Phys.\ Rev.\ D {\bf 71}, 123509 (2005)
[arXiv:hep-th/0504052];
M.~Sami, A.~Toporensky, P.~V.~Tretjakov and S.~Tsujikawa,
Phys.\ Lett.\ B {\bf 619}, 193 (2005);
G.~Calcagni, S.~Tsujikawa and M.~Sami,
Class.\ Quant.\ Grav.\ {\bf 22}, 3977 (2005)
[arXiv:hep-th/0505193];
 G.~Calcagni, B.~de Carlos and A.~De Felice,
 Nucl.\ Phys.\ B {\bf 752}, 404 (2006)
 [arXiv:hep-th/0604201];
B.~M.~Leith and I.~P.~Neupane,
JCAP {\bf 0705}, 019 (2007);
S.~Nojiri, S.~D.~Odintsov and P.~V.~Tretyakov,
 Phys.\ Lett.\ B {\bf 651}, 224 (2007)
 [arXiv:0704.2520 [hep-th]];
S.~Nojiri, S.~D.~Odintsov and M.~Sami,
Phys.\ Rev.\ D {\bf 74}, 046004 (2006)
[arXiv:hep-th/0605039];
B.~M.~N.~Carter and I.~P.~Neupane,
JCAP {\bf 0606}, 004 (2006)
[arXiv:hep-th/0512262];
T.~Koivisto and D.~F.~Mota,
Phys.\ Rev.\ D {\bf 75}, 023518 (2007)
[arXiv:hep-th/0609155];
G.~Cognola, E.~Elizalde, S.~Nojiri, S.~D.~Odintsov and S.~Zerbini,
Phys.\ Rev.\ D {\bf 75}, 086002 (2007)
[arXiv:hep-th/0611198];
M.~R.~Setare and E.~N.~Saridakis,
Phys.\ Lett.\ B {\bf 670}, 1 (2008);
A.~K.~Sanyal,
arXiv:0710.2450 [astro-ph];
B.~C.~Paul and S.~Ghose,
arXiv:0809.4131 [hep-th];
E.~Elizalde, S.~Jhingan, S.~Nojiri, S.~D.~Odintsov, M.~Sami and I.~Thongkool,
 Eur.\ Phys.\ J.\ C {\bf 53}, 447 (2008)
 [arXiv:0705.1211 [hep-th]];
A.~De Felice and T.~Suyama,
 arXiv:0904.2092 [astro-ph.CO].

\bibitem{Ito:2009nk}
Y.~Ito and S.~Nojiri,
arXiv:0904.0367 [hep-th].

\bibitem{Chame}
J.~Khoury and A.~Weltman,
Phys.\ Rev.\ Lett.\ {\bf 93}, 171104 (2004)
[arXiv:astro-ph/0309300].

\bibitem{Mota:2003tc}
D.~F.~Mota and J.~D.~Barrow,
Phys.\ Lett.\ B {\bf 581}, 141 (2004)
[arXiv:astro-ph/0306047].

\bibitem{thomas}
T.~P.~Sotiriou, V.~Faraoni and S.~Liberati,
 Int.\ J.\ Mod.\ Phys.\ D {\bf 17}, 399 (2008)
 [arXiv:0707.2748 [gr-qc]].

\bibitem{Hu:2007nk}
W.~Hu and I.~Sawicki,
Phys.\ Rev.\ D {\bf 76}, 064004 (2007).

\bibitem{acc}
S.~Nojiri and S.~D.~Odintsov,
Phys.\ Lett.\ B {\bf 657}, 238 (2007)
[arXiv:0707.1941 [hep-th]];
S.~Nojiri and S.~D.~Odintsov,
Phys.\ Rev.\ D {\bf 77}, 026007 (2008)
[arXiv:0710.1738 [hep-th]];
G.~Cognola, E.~Elizalde, S.~Nojiri, S.~D.~Odintsov, L.~Sebastiani and S.~Zerbini,
\textit{ibid}.\ {\bf 77}, 046009 (2008);
[arXiv:0712.4017 [hep-th]].

\bibitem{acc1}
S.~A.~Appleby and R.~A.~Battye,
Phys.\ Lett.\ B {\bf 654}, 7 (2007);
L.~Pogosian and A.~Silvestri,
Phys.\ Rev.\ D {\bf 77}, 023503 (2008);
S.~Nojiri and S.~D.~Odintsov,
Phys.\ Lett.\ B {\bf 652}, 343 (2007)
[arXiv:0706.1378 [hep-th]];
S.~Capozziello and S.~Tsujikawa, Phys.\ Rev.\ D {\bf 77}, 107501 (2008);
S.~Tsujikawa, Phys.\ Rev.\ D {\bf 77}, 023507 (2008)
[arXiv:0709.1391 [astro-ph]].

\bibitem{buch}
I.~L.~Buchbinder, S.~D.~Odintsov and I.~L.~Shapiro,
{\em Effective action in quantum gravity}, IOP Publishing, Bristol, 1992.

\bibitem{Cognola:2005sg}
G.~Cognola and S.~Zerbini,
 J.\ Phys.\ A {\bf 39}, 6245 (2006)
 [arXiv:hep-th/0511233].

\bibitem{fara}
V.~Faraoni,
 Phys.\ Rev.\ D {\bf 72}, 124005 (2005)
 [arXiv:gr-qc/0511094].

\bibitem{monica}
G.~Cognola, M.~Gastaldi and S.~Zerbini,
 Int.\ J.\ Theor.\ Phys.\ {\bf 47}, 898 (2008)
 [arXiv:gr-qc/0701138].

\bibitem{Cognola:2005de}
G.~Cognola, E.~Elizalde, S.~Nojiri, S.~D.~Odintsov and S.~Zerbini,
JCAP {\bf 0502}, 010 (2005)
[arXiv:hep-th/0501096].

\bibitem{frad}
E.~S.~Fradkin and A.~A.~Tseytlin,
Nucl.\ Phys.\ B {\bf 234}, (1984) 472.

\bibitem{Elizalde:1997nd}
E.~Elizalde, L.~Vanzo and S.~Zerbini,
Commun.\ Math.\ Phys.\ {\bf 194} (1998) 613
[arXiv:hep-th/9701060].

\bibitem{eli94}
E.~Elizalde, S.~D.~Odintsov, A.~Romeo, A.~A.~Bytsenko and S.~Zerbini
{\em Zeta regularization techniques with applications},
World Scientific, 1994;
E.~Elizalde,
{\em Ten physical applications of spectral zeta functions},
Lecture Notes in Physics, Springer-Verlag, Berlin, 1995;
A.~A.~Bytsenko, G.~Cognola, L.~Vanzo and S.~Zerbini,
Phys.\ Rept.\ {\bf 266} (1996) 1
[arXiv:hep-th/9505061];
A.~A.~Bytsenko, G.~Cognola, E.~Elizalde, V.~Moretti and S.~Zerbini,
{\em Analytic aspects of quantum fields}, World Scientific, Singapore, 2003.

\bibitem{tsamis}
N.~C.~Tsamis and R.~P.~Woodard,
 Class.\ Quant.\ Grav.\ {\bf 26}, 105006 (2009)
 [arXiv:0807.5006 [gr-qc]].

\bibitem{machado}
D.~Benedetti, P.~F.~Machado and F.~Saueressig,
 arXiv:0902.4630 [hep-th];
L.~N.~Granda and S.~D.~Odintsov,
 Phys.\ Lett.\ B {\bf 409}, 206 (1997)
 [arXiv:hep-th/9706062].

\bibitem{percacci}
A.~Codello, R.~Percacci and C.~Rahmede,
 Annals Phys.\ {\bf 324}, 414 (2009)
 [arXiv:0805.2909 [hep-th]].

\end{thebibliography}
\end{document}